# Dynamic control of catalysis within biological cells

**Biman Jana and Biman Bagchi**

Solid state and structural chemistry unit, Indian institute of science, Bangalore-12, India

E-mail:bbagchi@sscu.iisc.ernet.in

***Abstract***

**We develop a theory of enzyme catalysis within biological cells where the substrate concentration [S](t) is time dependent, in contrast to the Michaelis-Menten theory that assumes a steady state. We find that the time varying concentration can combine, in a non-linear way, with the ruggedness of the free energy landscape of enzymes (discovered both in single molecule studies and in simulations) to provide a highly efficient switch (or, bifurcation) between two catalytically active states, at a critical substrate concentration. This allows a dynamic control of product synthesis in cell.**

The concentrations of the constituents (Proteins, ATP, ADP, pH, etc.) in a healthy cell are strictly controlled and this control is dynamic in nature. The survival of the cell is crucially dependent on the several switches which operate with surprising efficiency in turning off and on the synthetic machines in the cell [1]. Synthesis of constituents in a cell is conducted by enzymes, sometime called “molecular machines”. A specific enzyme converts a particular substrate to a particular product by means of a specific chemical transformation [2]. Moreover, several studies have revealed that substrates are produced within a cell in a burst [3]. This may cause a dangerous imbalance in cell as concentration of that particular species can become alarmingly large subsequent to the burst. Enzyme should then start consuming that substrate at a high rate to bring back the concentration close to its physiological range. Once the imbalance is corrected, the machine should consume the substrate at a steady rate to prevent significant lowering of the concentration. A particular enzyme may thus need to act as a *substrate concentration dependent switch* that operates around a critical concentration of a substrate (physiological concentration). *However, how an enzyme performs its amazing task is still shrouded in mystery.*

A possible scenario is that an enzyme exists in two different conformational states below and above this critical concentration [4]. A biological cell must then have a sensor of substrate concentration that triggers a switch between the two states. Usual description of switch in cells employs multiple enzyme receptor sites and switching between different states occurs by receptor binding [5]. Nonlinear mathematics (such as bifurcation theory) can produce the essential characteristics of such switches [6-8]. For example, in the case of genetic toggle switch in Escherichia coli it is the receptor concentration that acts as the control parameter [5]. We are not aware of any discussion where substrate and/or product concentration act as the control parameters for an enzymatic switch.

Our understanding of the substrate concentration dependence of the enzyme kinetics is based on the Michaelis-Menten (MM) theory that assumes a time independent steady state substrate concentration [9]. Several theoretical studies have examined the validity of MM kinetics in the single molecular level [10-13]. The steady state operates under a chemical potential gradient where the in flow of the substrate and the out flow the product are guaranteed. Additionally, the substrate concentration is higher than that of the enzyme. Although the MM kinetics has been amply verified for *in vitro* experiments both at ensemble and at single molecular level [14], its validity inside the cell remains doubtful because inside the cell substrate concentration is time dependent and also the concentrations of enzyme and substrate are comparable.

The relevance of the above issues was made apparent in a recent experiment that revealed that GroEl enzyme that hydrolyzes ATP to ADP works in a different catalytic cycle when the concentration of ATP is high [4]. It was also observed that enzyme did not return to its equilibrium relaxed state after the product (ADP) was released. Instead, it was found to bind to ATP from an *intermediate state to continue the reaction cycle with a higher rate*. In addition, when ATP concentration became low, then the enzyme relaxed back to the equilibrium R-state where it was found to bind to ATP to continue the reaction cycle with a *lower rate* [4].

Recent detailed theoretical and very long computational study (~ 1μs) on the catalytic conversion of ATP and AMP to ADP by adenylate kinase (ADK) revealed the existence of a half-open half-closed (HOHC) *intermediate state* that can modify the catalytic cycle and accelerate the rate [15]. The intermediate state in an enzyme catalysis bears the similarity with the intermediate state in organic synthesis where the sequence of intermediates dictates the course of the reaction. Existence of such intermediate states in the free energy surface suggests ruggedness of the free energy surface. Some of these ruggedness arises due to water mediated interaction [15]. The intermediate HOHC state was conjectured to facilitate the catalysis by establishing a non-equilibrium steady state [16].

A recent theory of enzyme catalysis envisaged a non-equilibrium conformational cycle in which the active enzyme never needs to return to its native state during the cycle [17-18]. The cycle, which is maintained by steady inflow and outflow of the substrate and product, respectively, is driven of course by the free energy gradient. In this theory, depending on the rate of relaxation of the enzyme conformation after product release, the enzyme can capture a new substrate from an intermediate state and thereby the rate of catalysis can increase. We show below that with suitable generalization, such as the inclusion of the ruggedness of the free energy surface, one can have a molecular level description of a switch where the said mathematical non-linearity arises naturally.

We employ a free energy surface for enzyme catalysis as shown in **Figure 1**. The reaction cycle consists of several sequential steps. (i) The initiation of the reaction cycle which could be activated, (ii) the use of substrate-enzyme interaction to steer both towards reactive geometry, (iii) the catalytic conversion to form the product, (iv) the product release and subsequent enzyme relaxation back towards native state conformation, (v) the substrate capture and continuation of the cycle.

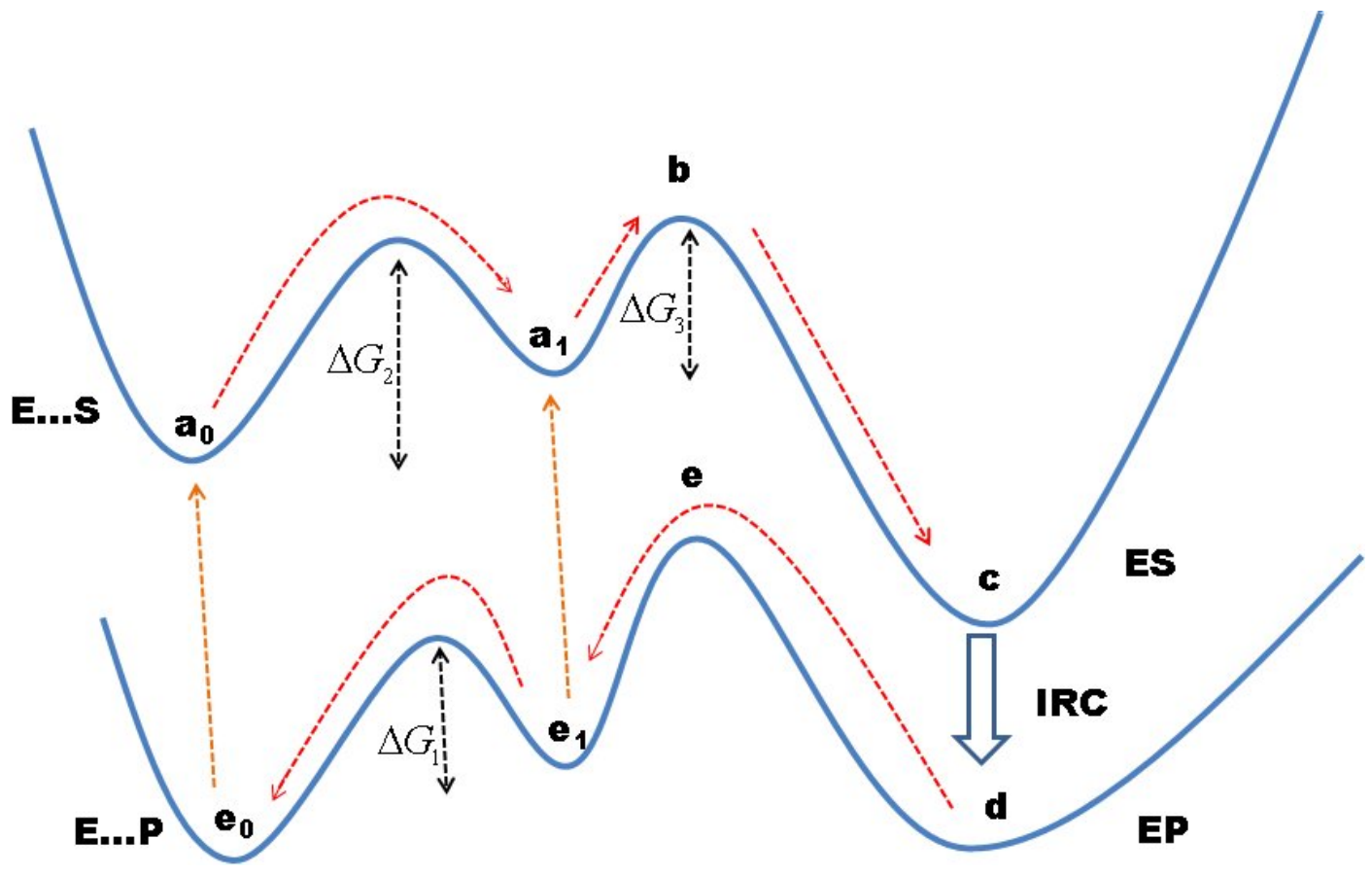

**Figure 1:** The free energy surface of enzyme catalysis. Intermediate state $e_1$ is introduced in the enzyme relaxation surface after the product release. A similar intermediate state $a_1$ is also introduced in the E…S similar to the $e_1$. The usual equilibrium states, $e_0$ and $a_0$, are also present in the present model. Relevant free energy barriers and the reaction steps are shown by arrows. Here $\Delta G_1$ determines the residence time of the $e_1$ state and also the substrate concentration dependent switch of the catalysis. The value of $\Delta G_1$ also determines the extent of ruggedness in the free energy surface of enzyme fluctuations.

We introduce an intermediate state ($e_1$) in the enzyme relaxation surface of the product side (EP surface) in addition to the equilibrium state ($e_0$), and an intermediate state ($a_1$) in addition to the conventional equilibrium state ($a_0$) of the substrate bound enzyme on the substrate site (ES surface). We have introduced the ruggedness in the free energy surface by introducing such intermediates in the spirit of our recent finding of new HOHC intermediate for the adenylate kinase enzyme [15, 16]. Introduction of such an intermediate can modify the catalytic cycle in the following manner: (1) while relaxing back to $e_0$ after the product release (e state), the enzyme now gets trapped in the $e_1$ state. The residence time of the enzyme in that minimum is dependent on the barrier ($\Delta G_1$) it experiences in the process of relaxation back to $e_0$. Now, if substrate encounters the $e_1$ state within its residence time, it goes to $a_1$ state directly in the ES surface. From the $a_1$ state it has to surmount a smaller barrier ($\Delta G_3$) to reach the downhill induced fit surface which takes the enzyme substrate complex to the reactive geometry. Clearly, this situation can happen at the high substrate concentration limit. The rate of catalysis for such a truncated cycling is high. (2) On the other hand, if substrate does not encounter the $e_1$ state within its residence time, it goes back to $e_0$ state and captures substrate to reach $a_0$ state of the ES surface. To reach the downhill induced fit surface, the $a_0$ state has to surmount two consecutive barriers ($\Delta G_2$ and $\Delta G_3$ ). Thus, in such a scenario of extended cycling, the reaction rate can become quite slow. Clearly, such a situation appears when the substrate concentration is low. Thus, in our present model we have a crossover between the truncated and extended cycling depending on the substrate concentration and the residence time of the enzyme in the intermediate state $e_1$.

According to the scheme shown in **Figure 1**, we can write the rate equations for different species. Several formalisms have been developed and discussed about motion on a rugged energy landscape [19, 20]. The rate equation for the state $e_0$ can be written as, $\frac{dp_{e_0}(t)}{dt} = k_{e_1 \to e_0}(\Delta G_1)\, p_{e_1}(t) - k_{e_0 \to a_0}([S])\, p_{e_0}(t)$. However, after the initiation of the reaction, this step can be pre-empted by the establishment of a non-equilibrium steady state that occurs at high substrate concentration, as described above and in the discussion of **Figure 1**. We show below that such a crossover indeed happens above a certain critical concentration [$S_C$] whose expression is provided later. This feature can be captured by introducing a Heaviside function (H) in the rate equations as,

$$\frac{dp_{e_0}(t)}{dt} = k_{e_i \to e_0}(\Delta G_1)\, H\left(k_{e_i \to e_0}(\Delta G_1) - k_{e_i \to a_1}([S])\right) p_{e_1}(t) - k_{e_0 \to a_0}([S])\, H\left(k_{e_i \to e_0}(\Delta G_1) - k_{e_i \to a_1}([S])\right) p_{e_0}(t)$$

$$\frac{dp_{e_1}(t)}{dt} = k_{\phi}\, p_{a_1}(t) - k_{e_i \to e_0}(\Delta G_1)\, H\left(k_{e_i \to e_0}(\Delta G_1) - k_{e_i \to a_1}([S])\right) p_{e_1}(t) - k_{e_1 \to a_1}([S])\left[1 - H\left(k_{e_i \to e_0}(\Delta G_1) - k_{e_i \to a_1}([S])\right)\right] p_{e_1}(t)$$

………….. (1)

Here $p_i(t)$ are the time dependent probability of species i and $k_{i \to j}$ are the rate of conversion from state i to j. Similar equations are also employed for $p_{a1}(t)$ and $p_{a0}(t)$. $k_{\varphi}$ is given by

$$\frac{1}{k_{\phi}} = \frac{1}{k_{a_1 \to b}} + \frac{1}{k_{b \to c}} + \frac{1}{k_{c \to d}} + \frac{1}{k_{d \to e_1}}$$

…………(2)

The mean first passage time (which is the inverse of the rate of the enzyme catalysis (ν)) can be obtained by solving the above equations as,

$$\frac{1}{v([S])}=\left[\frac{1}{k_{e_1\to a_1}([S])}+\frac{1}{k_\phi}\right]\left[1-H\left(k_{e_1\to a_0}(\Delta G_1)-k_{e_1\to a_1}([S])\right)\right]$$
$$+\left[\frac{1}{k_{e_0\to a_0}([S])}+\frac{1}{k_{e_1\to a_0}(\Delta G_1)}+\frac{1}{k_{a_0\to a_1}}+\frac{1}{k_\phi}\right]H\left(k_{e_1\to a_0}(\Delta G_1)-k_{e_1\to a_1}([S])\right)$$

………………..(3)

The first part of the left hand side of the above equation provides the rate when substrate concentration is higher and the second part will contribute when the substrate concentration is low. In Eq.3, [S] is an instantaneous substrate concentration. Note that Eq.3 follows essentially from the ruggedness of the free energy surface, shown in **Figure 1**, and from the kinetic scheme presented above. In the presence of time varying substrate concentration, it is not exact. Our numerical solution shows that Eq.3 remains accurate for the parameter space explored here. In Ref.18 a continuum version of essentially the same model was discussed. The results remain unchanged irrespective of the description used.

This solution provides a crossover between the two scenarios at a critical substrate concentration [$S_C$]. An analytical expression for this critical substrate concentration can be obtained from the above equations.

$$k_{e_1\to a_1}([S])|_{at\,[S_c]}=k_{e_1\to e_0}(\Delta G_1)$$
$$\Rightarrow [S_C]=\frac{k_{e_1\to e_0}(\Delta G_1)}{4\pi D_S a_s}=\frac{k^0_{e_1\to e_0}\exp(-\beta\Delta G_1)}{4\pi D_S a_s}$$

………………(4)

where we have used Smoluchowski reaction diffusion rate for the substrate capture step. $D_S$ and $a_S$ are the diffusion coefficient and the diameter of the substrate, respectively.

The substrate concentration dependent rates of catalysis for different values of $\Delta G_1$ are calculated using Eq 3. The results are shown in **Figure 2**. For very low value of $\Delta G_1$ (which corresponds to $[S_C] >> 10$) which implies a small residence time in the $e_1$ state, we find that substrate concentration dependence shows MM kinetics. From Eq. 4, small value of $\Delta G_1$ provides a very large value of $[S_C]$ which is practically not possible to attain (crossover is not observed in the concentration range shown here).

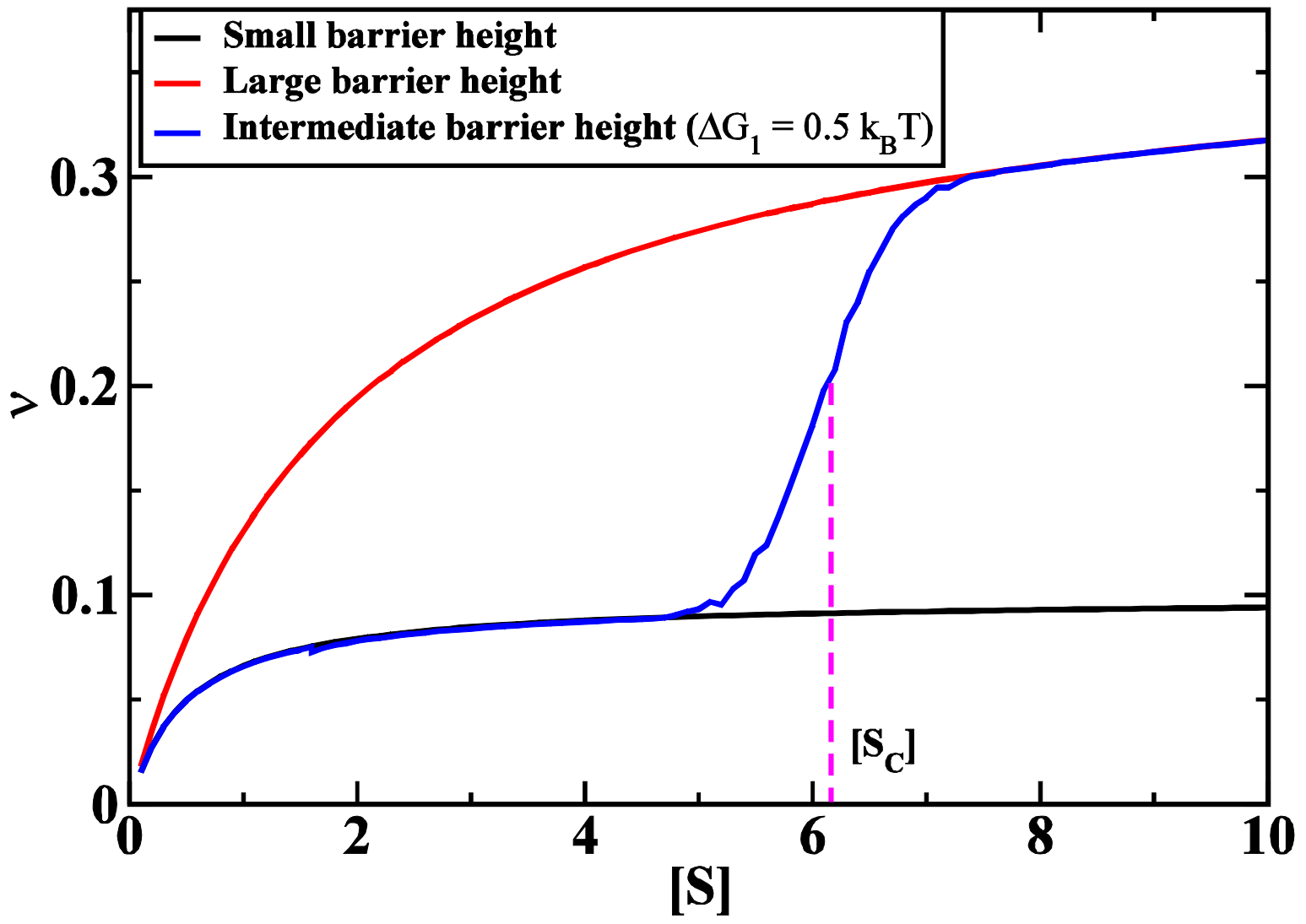

**Figure 2:** Substrate concentration dependence of the enzyme velocity (ν). Note the classical MM kinetics for the large and small barrier limit. Note also the presence of switch between the two types of MM behavior around critical substrate concentration for intermediate barrier height. We put $\frac{1}{4\pi D_s a_s} = 5.0$, $\frac{1}{k_{b\to c}} + \frac{1}{k_{c\to d}} + \frac{1}{k_{d\to e_1}} = 1$, $\frac{1}{k_{a_0\to a_1}} = \exp(2.0)$, $\frac{1}{k_{a_1\to b}} = \exp(0.5)$ and $\frac{1}{k_{e_1\to e_0}} = 0.1\exp(\Delta G_1)$. $\Delta G_1$ values are varied for different curves as discussed in the text.

On the other hand, for large value of $\Delta G_1$ (which corresponds to $[S_C]$ close to zero], the crossover substrate concentration is negligibly small for all practical purpose. Therefore, we again observe MM kinetics with large limiting enzyme velocity. Most interesting feature is observed when $\Delta G_1$ has intermediate value ($\Delta G_1$ = 0.5 $k_BT$). In such case, we observe a crossover from initial MM kinetics of small limiting enzyme velocity to another MM kinetics of large enzyme velocity at the higher substrate concentration side. The crossover is observed around the critical substrate concentration ($[S_c] \approx 6.2$) calculated from Eq. 4. We also calculate the substrate concentration dependent rate for different intermediate value of $\Delta G_1$ and plot in **Figure 3**. We find that all the plots show similar behavior as discussed above. However, the value of critical substrate concentrations for these plots varies. It increases with decrease in $\Delta G_1$ value. As we have discussed earlier, the value of $\Delta G_1$ provides a measure of the ruggedness in the free energy surface of enzyme fluctuations.

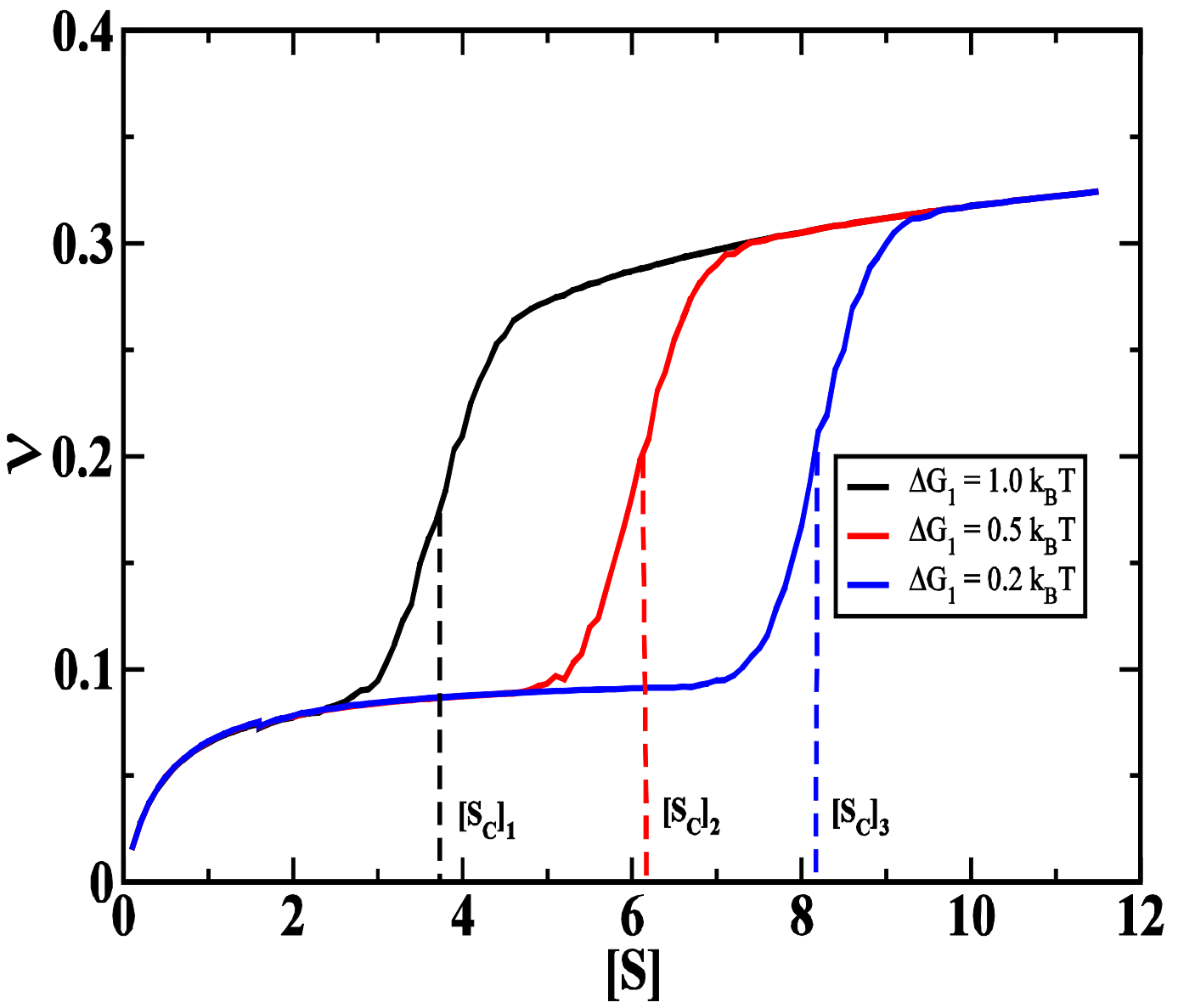

**Figure 3:** Switching behavior of the substrate concentration dependence of the enzyme catalysis rate for different value of ruggedness parameter. Ruggedness dictates the critical substrate concentration for the switch. As ruggedness increases critical substrate concentration decreases. We put values of all the parameters as mentioned in the caption of Figure 2. $\Delta G_1$ values are varied for different curves as shown in the figure legend.

Thus, ruggedness in the free energy surface dictates the critical substrate concentration around which switching of the enzyme function occurs. Our results suggest that if the concentration of a substrate in a healthy cell is small, the corresponding enzyme which consumes that substrate has a conformational free energy surface with large ruggedness and vice versa.

As stated earlier, in a biological cell, substrate is produced in burst. Therefore, the initial substrate concentration ($[S_0]$) can become high. This triggers the enzyme to work at a higher rate. Enzyme starts consuming the substrate according to Eq. 3 and as a result of that substrate concentration decreases. Thus, in cell substrate concentration is time dependent. In the rate equation (Eq. 3), both $k_{e_0 \to a_0}$ and $k_{e_1 \to a_1}$ are substrate capture rate and therefore dependent on the instantaneous substrate concentration. Rate also becomes time dependent in cell. Substrate concentration in cell follows $\left[S(t)\right] = \left[S_0\right] - \int_0^t dt' v(t')$.

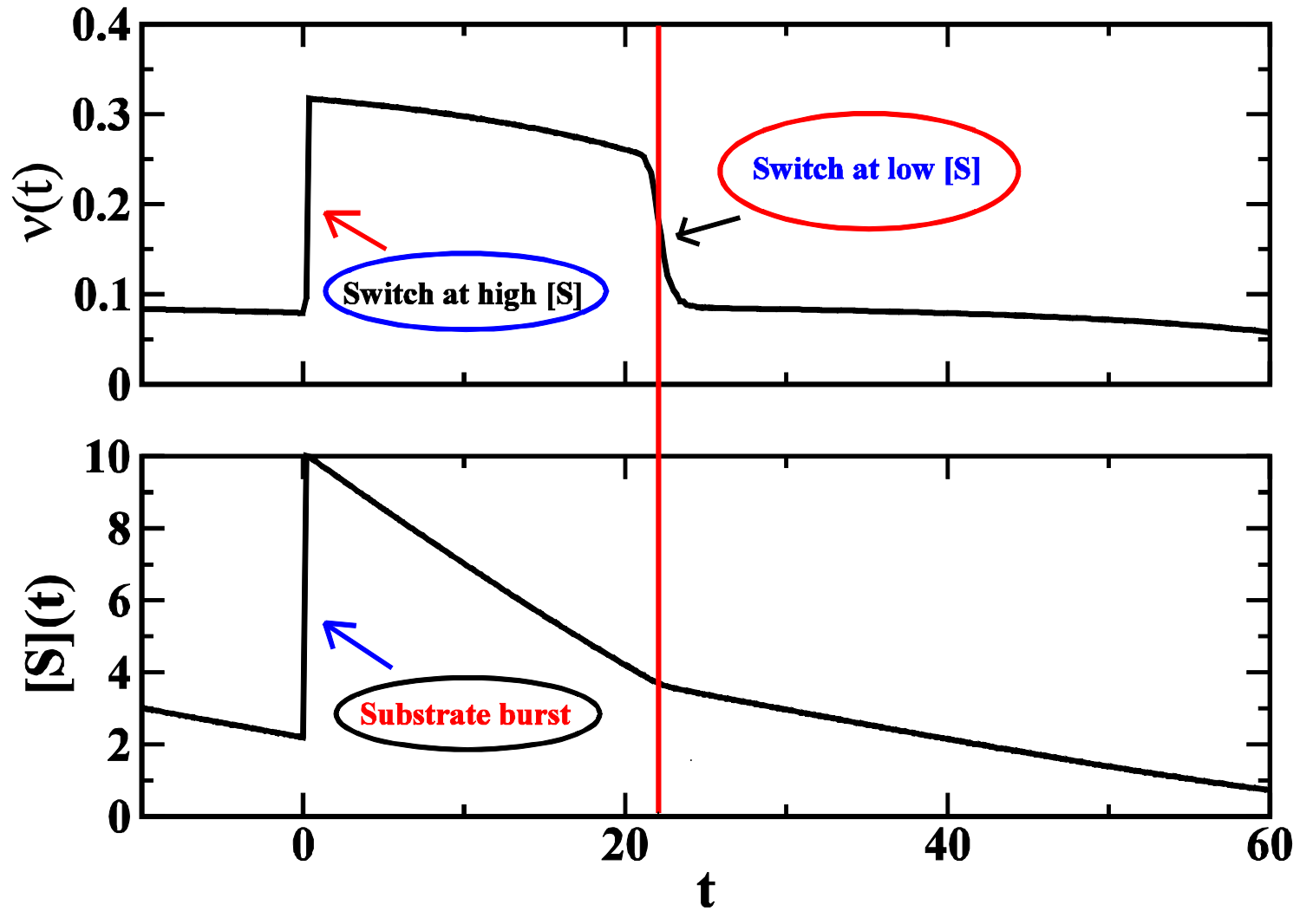

**Figure 4:** Time dependent enzyme velocity (ν(t)) and substrate concentration ([S](t)) within a biological cell. At t=0, the substrate produced in cell at burst which makes substrate concentration very high ($[S_0]$ = 10]. Note the switch of the enzyme kinetics around t= 22.0. Note also the large and small negative slope of the time dependence of substrate concentration decay across the switch subsequent to substrate burst. Note also the switch (at high substrate concentration) after the substrate burst (t = 0). We took $\Delta G_1 = 1.0\ k_BT$ for the calculation. The values of other parameters are same as mentioned in the caption of Figure 2

We now solve Eq. 3 with time dependent substrate concentration self consistently to get time dependent rate of the catalysis which is shown in **Figure 4(a)**. As the initial substrate concentration is high (subsequent to burst at t =0), the rate readily becomes high (switch at high [S]) and decreases with time slowly. When the substrate concentration reaches a value near the critical value (here, $[S_C] \approx 3.7$ in reduced unit for the parameters used), the rate suddenly decreases. This gives rise to a switch in the time dependence of the rate of catalysis. The above calculation also provides the time dependent substrate concentration which is shown in **Figure 4(b)**. [S](t) decreases rapidly (after the burst) till it reach its critical value and then it starts to decrease slowly. These results indicate the following: (1) when the substrate concentration becomes alarmingly high, enzyme starts to work at a high rate to consume the substrate faster [switch at high [S]). (2) When the substrate concentration reaches its physiological limit (critical concentration), enzyme starts to work at a lower rate (switch at low [S]) and the substrate concentration does not further change rapidly. Thus, such an enzymatic switch is a necessary mechanism for the cell to maintain the physiological concentration of its constituents.

While we are not aware of any previous theoretical study that discusses the possibility of a switch driven by time dependent substrate concentration, the results derived here from the theoretical model of dynamic control (and switch) are in good agreement with experiment models and simulation studies. Most importantly, our model offers perhaps the simplest example of the switch and is based on time varying substrate concentration: the switch is turned on when the concentration of the substrate is large. A similar switch can be built that is based on product concentration. When the product concentration becomes large, the product release step can become slow, thus again switching off the molecular machine. Thus, the mechanism proposed and developed here can be incorporated within a network within the cell where different molecular machines microscopically "talk" with each other through time varying substrate and product concentrations. In such a network, product from one machine is the substrate or fuel for a second machine, and so on.

The merit of the present model is that it reveals that fact how enzyme uses conformational landscape of its own to maintain a critical concentration of important substrates inside the cell. Although, the present model is also simple, it essentially describes the substrate concentration dependent switching of the enzymatic activity in a novel way.

**ACKNOWLEDGEMENT**

We thank DST (India) for partial support of this work. BB further acknowledges support through a JC Bose Fellowship from DST.